\documentclass[journal,twoside]{IEEEtran}
\usepackage{graphicx,amsmath,amssymb}
\usepackage{verbatim}
\usepackage[noadjust]{cite}
\usepackage{array}
\usepackage{cases}
\usepackage{url}
\usepackage{multirow}
\usepackage{cleveref}
\newcommand{\bra}[1]{\left\vert #1 \right\rangle}
\newcommand{\ket}[1]{\left\langle #1 \right\vert}

\newtheorem{theorem}{Theorem}[section]
\newtheorem{proposition}[theorem]{Proposition}

\newtheorem{problem}[theorem]{Problem}

\begin{document}
\title{Parsing a Sequence of Qubits}
\author{Yuichiro~Fujiwara,~\IEEEmembership{Member,~IEEE}%
\thanks{The author acknowledges support from JSPS Postdoctoral Fellowships for Research Abroad.}%
\thanks{Y. Fujiwara is with the Division of Physics, Mathematics and Astronomy, California Institute of Technology, MC 253-37, Pasadena, CA 91125 USA
(email: yuichiro.fujiwara@caltech.edu).}
\thanks{Copyright \copyright\ 2013 IEEE. Personal use of this material is permitted.
However, permission to use this material for any other purposes must be obtained from the IEEE by sending a request to pubs-permissions@ieee.org.}}
\markboth{IEEE transactions on Information Theory,~Vol.~x, No.~xx,~month~year}
{Fujiwara: Parsing a sequence of qubits}

%\pubid{0000--0000/00\$00.00~\copyright~2007 IEEE}

\maketitle

\begin{abstract}
We develop a theoretical framework for frame synchronization, also known as block synchronization, in the quantum domain which
makes it possible to attach classical and quantum metadata to quantum information over a noisy channel
even when the information source and sink are frame-wise asynchronous.
This eliminates the need of frame synchronization at the hardware level
and allows for parsing qubit sequences during quantum information processing.
Our framework exploits binary constant-weight codes that are self-synchronizing.
Possible applications may include asynchronous quantum communication such as
a self-synchronizing quantum network where one can hop into the channel at any time,
catch the next coming quantum information with a label indicating the sender,
and reply by routing her quantum information with control qubits for quantum switches all without assuming prior frame synchronization between users.
\end{abstract}

\IEEEpeerreviewmaketitle

\maketitle

\section{Introduction}
Quantum information processing is expected to make various kinds of seemingly formidable task
manageable and may make what is impossible in the classical domain possible \cite{MikeandIke}.
In order to realize such large-scale quantum computation and quantum communication,
it is vital to develop theoretical foundations.
To this end, various notions have been introduced and investigated in quantum information science,
some of which are quantum analogues of important concepts in classical information theory
and others unique to the quantum world.

Frame synchronization is one of the most important building blocks of basic data manipulations.
Because virtually all information we process has some kind of frame structure,
without synchronization between the information source and sink, one can not manipulate information as intended.
For instance, the phrase ``classical information'' will lose its meaning if we combine the two words into
one string of twenty letters ``classicalinformation'' or divide it at a different position as in ``classicalin formation.''
The English text the reader is reading now is a sequence of blocks of classical information, or words,
with spaces as metadata that indicate the boundaries between each information block.

A typical example in a communications scenario is \textit{packet framing} for a classical communication network,
where the sender encapsulates each packet of data into a frame
and provides information the network needs to deliver the data, such as destination addresses.
The additional information is metadata, or \textit{overhead data},
to the actual information content, or the \textit{payload}.
The supplementary data are often provided through a set of extra bits appended to the payload, called a \textit{packet header} or \textit{trailer},
depending on where they are placed.
The simplest but important use of such metadata is to indicate boundaries between frames, that is, framing data into units \cite{ComputerNetworks}.

Looking at impressive promises quantum information science makes, such as exponential speedups of particular kinds of computation \cite{Mosca},
one might assume that attaching metadata to quantum information must be as easy as in the classical domain.
This is not the case, however.

To illustrate the challenge in the quantum domain, take the simplest metadata as an example,
namely the boundary signal between each block of information.
In the classical domain, the simplest way to frame data is to insert a marker at the start or end of each information block
as in spaces between words in the English language.
If data are represented by binary sequences of zeros and ones,
a fixed binary pattern may be used as a special codeword to signal each boundary.

In the quantum domain, things are not that straightforward.
Let us assume that we transmit a sequence of qubits carrying quantum information.
If we simply inserted a certain pattern of quantum states
$(\alpha_0\bra{0} + \beta_0\bra{1})\otimes\cdots\otimes(\alpha_s\bra{0} + \beta_s\bra{1})$
as a boundary signal between each information block,
the receiver would have to perform some measurement to pick up on the inserted signals.
Because measurement generally disturbs quantum information carried by qubits,
the receiver must perform the measurement exactly on the synchronization signals,
which is only possible if the receiver already knows where the boundaries are to begin with.

The only known theoretical solution to this catch-22 is quantum synchronizable coding \cite{Fblock,Fblock2}.
However, this is not packet framing for asynchronous quantum channels per se;
quantum synchronizable coding is a method for providing efficient recovering when frame synchronization may become slightly off.
In other words, it assumes prior establishment of frame synchronization before the start of data transmission and
corrects synchronization errors of small magnitude if they occur during the transmission.
All other solutions are hardware dependent in principle,
which confines the possibilities of quantum information processing to a limited realm.

The purpose of the present paper is to develop an information theoretic framework
for attaching metadata to quantum information to open paths to complex frame-wise asynchronous quantum information processing.
We assume no prior frame synchronization between the information source and sink,
so that all metadata including boundary information must be embedded in the transmitted data.
We also assume that the channel is noisy, so that all information must be protected from quantum noise.

Note that the concept of metadata is often quite ambiguous if we do not specify the context.
Our goal is to make the first step toward frame-wise asynchronous quantum communication and quantum computation
in a fashion similar to traditional packet framing for classical information.
Hence, we restrict the concept of metadata to the kind of information extra classical bits in a frame would typically carry in a frame-wise asynchronous situation.
Trivially, overhead data of this kind can generally be attached through extra qubits in a frame if we solve the problem of locating boundaries between frames.
Hence, in what follows, we mostly focus on frame synchronization.

In Section \ref{model} we mathematically describe frame synchronization in the quantum setting
and present a naive approach that mimics the straightforward method of inserting spaces between words.
This will be generalized to develop a more general framework for parsing a sequence of qubits in Section \ref{framework}.
Concluding remarks are given in Section \ref{conclude}.

\section{Mathematical model}\label{model}
We follow the simple model introduced for quantum synchronizable codes \cite{Fblock}.
Let $Q = (q_0,\dots, q_{l-1})$ be an ordered set of length $l$, where each element represents a qubit.
A \textit{frame} $F_i$ is a set of consecutive elements of $Q$.
Let $\mathcal{F} = \{F_0, \dots, F_{m-1}\}$ be a set of frames.
The ordered set $(Q, \mathcal{F})$ is called a \textit{framed sequence} if
the elements of a sequence are partitioned into groups of consecutive elements, that is,
$\vert\bigcup_i F_i\vert = l$ and $F_i \cap F_j = \emptyset$ for $i \not= j$.

Take a set $G = \{q_j, \dots, q_{j+g-1}\}$ of $g$ consecutive elements of $Q$.
We call $G$ the \textit{window}.
The window $G$ is \textit{misaligned} by $r$ qubits to the \textit{right} with respect to $(Q, \mathcal{F})$
if there exits an integer $r$ and a frame $F_i$ such that $F_i = \{q_{j-r}, \dots, q_{j+g-r-1}\}$ and $G \not\in \mathcal{F}$.
If $r$ is negative, we may say that $G$ is misaligned by $\vert r \vert$ qubits to the \textit{left}.
$G$ is \textit{properly aligned} if $G \in \mathcal{F}$.

As a concrete example, take three qubits and encode each qubit into nine qubits by Shor's nine qubit code \cite{Shor}.
The resulting 27 qubits may be seen as $Q = (q_0, \dots, q_{26})$, where the three encoded nine qubit blocks
$\left\vert \varphi_0 \right\rangle$, $\left\vert \varphi_1 \right\rangle$, and $\left\vert \varphi_2 \right\rangle$ form frames
$F_0 = (q_0, \dots, q_8)$, $F_1 = (q_9, \dots, q_{17})$, and $F_2 = (q_{18}, \dots, q_{26})$ respectively.
We assume that a device knows that the size of each information block is nine.
If misalignment occurs by, say, two qubits to the left,
the device that tries to correct errors on qubits in $\left\vert \varphi_1 \right\rangle$ applies the error correction procedure
to the window $G$ of nine qubits $q_7$, \dots, $q_{15}$, two of which come from $F_0$ and seven of which $F_1$.
In this case, when measuring the stabilizer generator $IZZIIIIII$ of the nine qubit code to obtain the syndrome,
what the device actually does to the whole system can be expressed as
\[I^{\otimes 8}ZZI^{\otimes17}\left\vert \varphi_0 \right\rangle\left\vert \varphi_1 \right\rangle\left\vert \varphi_2 \right\rangle,\]
which, if frame synchronization were correct, would be
\[I^{\otimes 10}ZZI^{\otimes15}\left\vert \varphi_0 \right\rangle\left\vert \varphi_1 \right\rangle\left\vert \varphi_2 \right\rangle.\]
$I^{\otimes 8}Z$ does not stabilize $\left\vert \varphi_0 \right\rangle$, nor does $ZI^{\otimes 8}$ $\left\vert \varphi_1 \right\rangle$.
Hence, errors are introduced to the system, rather than detected or corrected.

The objective of frame synchronization is to ensure that the device can identify the location of each boundary
and that the window on which quantum operations are performed by the device is always properly aligned.
Note that if we assume that the magnitude of misalignment can be more than the shortest frame length,
it may happen that the window is ``properly aligned'' to an unintended frame.
However, in principle, proper alignment ensures synchronization recovery
if each frame is labeled by metadata in some form of classical information such as an integer.
Hence, we first focus on proper alignment and then consider more general situations.

In the remainder of this section, we present a naive approach to frame synchronization to illustrate one of our key ideas.
To describe a straightforward quantum analogue of inserting spaces between words,
we employ a Hilbert space of dimension three where
we have three orthogonal basis states $\bra{0}$, $\bra{1}$, and $\bra{2}$.
For computation, we work on a tensor product of $2$-dimensional Hilbert spaces contained as a subspace of the system space,
so that quantum information is processed by using $\bra{0}$ and $\bra{1}$ in the same way as in an ordinary qubit system.
The third basis state $\bra{2}$ is allocated for synchronization.

In order for our framework to be as general as possible,
we do not specify the actual implementation method of the multilevel quantum system.
This is also because we require little more than what is already assumed in an ordinary qubit system over the quantum erasure channel.
For the basics of coding theory for the quantum erasure channel and a brief review of multilevel quantum systems,
the reader is referred to \cite{GBP,LGZZYP,GT} and references therein.
Similarly, we do not consider a peculiar type of synchronization error specific to a particular physical implementation.
For such specialized models in the classical domain, we refer the reader to \cite{MBT,MTL}.

Assume that we have a sequence of qutrits, where quantum information is carried by qubits lying in the 2-dimensional subspaces
and the state $\bra{2}$ is inserted as a synchronization signal between each information block.
We perform the measurement $\{\bra{0}\ket{0} + \bra{1}\ket{1},  \bra{2}\ket{2}\}$ on each qutrit.
Because our qubits are orthogonal to $\bra{2}$ and will not be affected by $\bra{0}\ket{0} + \bra{1}\ket{1}$,
in principle we can deterministically identify the locations of $\bra{2}$s without disturbing the quantum states.
For instance, if the state $\bra{2}$ is used only at the beginning of each information block,
we can successfully locate the boundaries as desired.

If we can assume that there is no quantum noise, this simple approach will work just fine.
However, such an assumption is unreasonable regardless of how a quantum system is implemented.
In fact, it is easy to see that this method can not even tolerate a single erroneous qutrit.
For instance, it is reasonable to assume
that the post-measurement state of what used to be a qubit in the 2-dimensional subspace may end up
outside of the subspace due to quantum noise as in the quantum erasure channel.
In this case, even a single quantum error can trick a synchronization device into recognizing the wrong position in the middle of an information block as a boundary.

\section{Quantum tags for error tolerance}\label{framework}
In this section we present a generalized method for asynchronously attaching metadata under the presence of quantum noise.
In what follows, we assume that qutrits may be subjects of any quantum error due to noise,
so that they may result in any states, possibly entangled with others unintended ways.

This section is divided into four subsections.
Subsection \ref{s3a} provides basic notions employed throughout this paper.
Subsection \ref{s3b} is devoted to generalizing the native approach described in the previous section
in order to realize a simple error-tolerant synchronization scheme.
Further generalization is given in Subsection \ref{s3c}, where attaching supplementary classical information along with boundary information is discussed.
Subsection \ref{s3d} explores possible use of our methods as packet headers in a more general situation in which the lengths of frames may vary.

We will employ mathematical tools, notions, and some basic facts from
coding theory, combinatorial and algebraic design theory, finite geometry, sequences, and finite fields.
While we only assume familiarity with quantum information,
the reader who wishes more backgrounds and thorough treatments is referred to standard textbooks in respective fields such as
\cite{HP,BJL,Hirschfeld,FD,finitefield}.

\subsection{Preliminaries}\label{s3a}
As in the simple method we described in the previous section, we insert $\bra{2}$s as auxiliary markers.
The difference is that the inserted positions form a carefully designed pattern within each information block.
As before, the system space is regarded as a tensor product $\mathcal{H}_{\text{sys}} = \mathcal{H}_3\otimes\cdots\otimes\mathcal{H}_3$
of $3$-dimensional Hilbert spaces, each of which contains a $2$-dimensional subspace that may be used for computing.
Hence, the subspace of computational states can be written as $\mathcal{H}_{\text{comp}}=\mathcal{H}_2\otimes\cdots\otimes\mathcal{H}_2$.
The system space is a direct sum $\mathcal{H}_{\text{sys}} = \mathcal{H}_{\text{comp}}\oplus\mathcal{H}_{\text{comp}}^{\perp}$ of subspaces.
In what follows, the symbol $\mathcal{H}_2$ always implies the $2$-dimensional subspace for computing spanned by $\bra{0}$ and $\bra{1}$.

As in the case of the naive approach,
the receiver performs the measurement $\{\bra{0}\ket{0} + \bra{1}\ket{1}, \bra{2}\ket{2}\}$ on each qutrit
in order to recognize the predetermined patterns of inserted $\bra{2}$s.
We use the finite field $\mathbb{F}_2$ of order two to represent each measurement outcome
by assigning $0$ and $1$ to the measurement results of the measurement operators $\bra{0}\ket{0} + \bra{1}\ket{1}$
and $\bra{2}\ket{2}$ respectively.
Hence, if the information sink performs the measurement on $v$ consecutive qutrits in a sequence of qutrits,
we obtain a binary $v$-dimensional vector as its \textit{outcome vector}.

If some qutrits are suffering from quantum errors, an outcome vector may also be altered.
In terms of measurement results, there are two types of error.
One is what used to be a qubit in $\mathcal{H}_{2}$ yielding measurement result $1$ due to a quantum error on it.
The other is a qutrit that was originally $\bra{2}$ giving the other measurement result.
By borrowing terminology from the quantum erasure channel literature,
the former kind is called an \textit{erasure}. We call the latter kind an \textit{incursion}.
Note that from the viewpoint of the receiver an erasure adds a measurement result $1$ and that an incursion makes one disappear.

For a frame $F = (q_0, \dots, q_{v-1})$ with some $\bra{2}$s in a framed sequence of qutrits,
define binary $v$-dimensional vector $\boldsymbol{x}_F = (x_0^{(F)}, \dots, x_{v-1}^{(F)})$ over $\mathbb{F}_2$
by \[x_i^{(F)} =
\begin{cases}
0 & \text{if} \ q_i \in \mathcal{H}_2,\\
1 & \text{otherwise}.
\end{cases}\]
We call this binary vector the \textit{tag vector} of $F$.

As usual, the \textit{support} $\text{supp}(\boldsymbol{x})$ of a $v$-dimensional vector $\boldsymbol{x} = (x_0, \dots, x_{v-1})$
over a finite field is the set $\{i \ \vert \ x_i \not= 0\}$ of coordinates in which the corresponding entries are nonzero.
Hence, given the tag vector $\boldsymbol{x}_F$ of a frame $F$,
the support $\text{supp}(\boldsymbol{x}_F)$ is simply the set of positions in which $\bra{2}$s are inserted.
We call this set the $(v,k)$-\textit{quantum tag} of \textit{length} $v$ and \textit{weight} $k$ attached to the frame $F$
if $\vert \text{supp}(\boldsymbol{x}_F) \vert = k$.\footnote{Despite its name,
a ``quantum'' tag will be used to provide metadata through preprocessing which is essentially classical.
In this sense, one may think of it as a ``classical'' tag that is designed for parsing a stream of quantum information.}
Fig.\ \ref{fig1} shows a frame $(q_0,\dots,q_{26})$ tagged by $\{0,3,11,21,26\}$.
\setlength{\unitlength}{8.4mm}
\begin{figure}
\centering
\begin{picture}(10,0.5)
\put(0,0){\line(1,0){9.8}}
\put(0,0.5){\line(1,0){9.8}}
\put(0.1,0){\line(0,1){0.5}}
\put(0.4,0){\line(0,1){0.5}}
\put(0.7,0){\line(0,1){0.5}}
\put(1.0,0){\line(0,1){0.5}}
\put(1.3,0){\line(0,1){0.5}}
\put(1.6,0){\line(0,1){0.5}}
\put(1.9,0){\line(0,1){0.5}}
\put(2.2,0){\line(0,1){0.5}}
\put(2.5,0){\line(0,1){0.5}}
\put(2.8,0){\line(0,1){0.5}}
\put(3.1,0){\line(0,1){0.5}}
\put(3.4,0){\line(0,1){0.5}}
\put(3.7,0){\line(0,1){0.5}}
\put(4,0){\line(0,1){0.5}}
\put(4.3,0){\line(0,1){0.5}}
\put(4.6,0){\line(0,1){0.5}}
\put(4.9,0){\line(0,1){0.5}}
\put(5.2,0){\line(0,1){0.5}}
\put(5.5,0){\line(0,1){0.5}}
\put(5.8,0){\line(0,1){0.5}}
\put(6.1,0){\line(0,1){0.5}}
\put(6.4,0){\line(0,1){0.5}}
\put(6.7,0){\line(0,1){0.5}}
\put(7,0){\line(0,1){0.5}}
\put(7.3,0){\line(0,1){0.5}}
\put(7.6,0){\line(0,1){0.5}}
\put(7.9,0){\line(0,1){0.5}}
\put(8.2,0){\line(0,1){0.5}}
\put(8.5,0){\line(0,1){0.5}}
\put(8.8,0){\line(0,1){0.5}}
\put(9.1,0){\line(0,1){0.5}}
\put(9.4,0){\line(0,1){0.5}}
\put(9.7,0){\line(0,1){0.5}}

\put(0.4,0){\rule{2.52mm}{4.2mm}}
\put(1.3,0){\rule{2.52mm}{4.2mm}}
\put(3.7,0){\rule{2.52mm}{4.2mm}}
\put(6.7,0){\rule{2.52mm}{4.2mm}}
\put(8.2,0){\rule{2.52mm}{4.2mm}}
\end{picture}
$\underbrace{\phantom{000000000000000000000000000000llllllllllllll}}_\text{Frame}$ \phantom{illlllll}
\caption{Frame with a $(27,5)$-quantum tag. $\vert$
The black boxes are $\bra{2}$s. The white boxes are qubits carrying the payload quantum information.
The frame of size $27$ is tagged by $\{0,3,11,21,26\}$.
Interpreting white boxes as $0$s and block boxes $1$s gives the corresponding tag vector.}\label{fig1}
\end{figure}
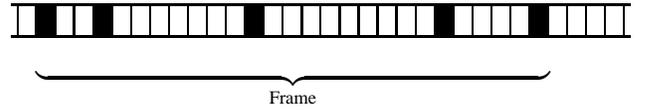
We may speak of the tag vector of a quantum tag to mean its binary vector representation when it is not important to specify a frame.

Define bijection $\pi$ from $\mathbb{F}_2^n$ onto $\mathbb{F}_2^n$ itself
as the cyclic shift by one bit to the right, that is,
$\pi$ maps a $v$-dimensional vector $\boldsymbol{x} = (x_0, \dots, x_{v-1})$ to $\pi(\boldsymbol{x}) = (x_{v-1},x_0,\dots,x_{v-2})$.
We also define the set version of cyclic shifts.
For a subset $S \subseteq \textit{\textbf{Z}}_v$ of the ring of integers modulo $v$ and $i \in \textit{\textbf{Z}}_v$,
the \textit{translate} of $S$ by $i$ is defined as $S + i = \{s + i \pmod{v} \ \vert \ s \in S\}$.

\subsection{Optimal synchronization with one quantum tag}\label{s3b}
Here we present a method for inserting marker qutrits in an error-tolerant fashion.
We assume that $k$ $\bra{2}$s are inserted at the information source to form a tagged frame.
We also assume that every frame is of the same size $v$, so that
the payload of a frame consists of $v - k$ qutrits in $\mathcal{H}_2$, or in other words, $v - k$ qubits.

This subsection studies what kind of quantum tag $T$ gives the best error tolerance for given $v$ and $k$.
It will be shown that this problem is exactly that of designing a binary sequence of good periodic auto-correlation properties \cite{FD}
while keeping in mind the peculiar constraint that
the number of $1$s in the sequence directly affects the information rate in our quantum information processing scenario.
With a simple argument,
we will prove that the best binary sequences that have been studied in the classical sequence design literature without this constraint
are still optimal for our purpose.

We protect the payload by a conventional quantum error-correcting code such as stabilizer codes \cite{Gdissertation}.
Note that once frame synchronization is achieved,
the receiver can safely perform a quantum error correction procedure on the payload.
Thus, any quantum error can be treated as would be in a standard quantum error correction scheme.
For instance, if some qubits fall to the outside of the $2$-dimensional subspaces for computing,
they can be corrected by the binary quantum error-correcting code used for the payload by regarding them as located errors
and manually setting erroneous qutrits to an arbitrary qubit in $\mathcal{H}_{2}$ \cite{GBP}.
Quantum errors that keep qubits within $\mathcal{H}_{\text{comp}}$ and a combination of the two types of error can be corrected
straightforwardly by the chosen quantum error-correcting code as long as the number of such errors does not exceed the error correction capacity of the code.
Quantum errors on tags can also be corrected, for example, by manually recreating the pattern of $\bra{2}$s for each frame.

Thus, our objective is to develop a technique that allows for the receiver to correctly recognize quantum tag $T$
within a sequence of qutritis even if it is buried under quantum noise.
The field of signal processing provides mathematical tools to formalize the problem of recognizing a pattern of signals.
We employ the notion of discrete correlation to build a foundation of the theory of quantum tags that suit our needs.

The \textit{periodic autocorrelation} $R_{\boldsymbol{x}}(t)$ at \textit{time} $t$ of a $v$-dimensional vector
$\boldsymbol{x} = (x_0, \dots, x_{v-1})$ over $\mathbb{F}_2$ is defined as the inner product
\begin{align*}
R_{\boldsymbol{x}}(t) &= \boldsymbol{x}\cdot\pi^{t}(\boldsymbol{x})\\
&= \sum_{0 \leq i \leq v-1}x_ix_{i+t}
\end{align*}
between $\boldsymbol{x}$ and a cyclic shift of itself, where the subscripts in the sum are taken modulo $v$.
Trivially, $R_{\boldsymbol{x}}(0) = \vert \text{supp}(\boldsymbol{x})\vert$ for any $v$-dimensional vector $\boldsymbol{x}$.
The values  $R_{\boldsymbol{x}}(t)$ for $t \not\equiv 0 \pmod{v}$ are the \textit{off-peak} autocorrelations.
If $R_{\boldsymbol{x}}(t) =  \vert \text{supp}(\boldsymbol{x})\vert$ for some $t \not\equiv 0 \pmod{v}$,
it means that the orbit $Orb_{\pi}(\boldsymbol{x})$ of $\boldsymbol{x}$ over cyclic shifts is shorter than $v$.
In other words, cyclically shifting $\boldsymbol{x}$ by $t$ bits to the left results in the same vector again.

A $(v,k)$-quantum tag is \textit{self-synchronizing} if there exists a positive integer $\rho$
such that
\[1 \leq  \rho \leq 2\left(k - \max_{1 \leq t \leq v-1}\left(R_{\boldsymbol{x}}(t)\right)\right),\]
where $\boldsymbol{x}$ is the corresponding tag vector.
We may say that such a tag is a self-synchronizing $(v,k,\rho)$-quantum tag.
The parameter $\rho$ is the \textit{comma-free index}.
Note that, by definition, a self-synchronizing quantum tag of comma-free index $\rho > 1$ is also of comma-free index $\rho'$ for any positive integer $\rho' < \rho$.
Informally, the comma-free index is a measure of guaranteed dissimilarity between a tag vector and its cyclic shifts.
To put it in the set theoretic language, for any $t \not\equiv 0 \pmod{v}$ a self-synchronizing $(v,k,\rho)$-quantum tag $T$ and its translate $T+t$
have at least $\rho$ elements that belong to only one of them,
or equivalently,
\[\vert T \cup(T+t)\vert-\vert T \cap(T+t)\vert \geq \rho.\]
As a nontrivial example, take the set $\{0,3,11,21,26\} \subset \textit{\textbf{Z}}_{27}$ we used in Fig.\ \ref{fig1}.
It is routine to check that this is a self-synchronizing $(27,5,8)$-quantum tag.
For instance, $\{0,3,11,21,26\}$ and its translate $\{1,4,12,22,0\}$ only share the element $0$.
Hence, there are eight elements $\{3,11,21,26\}\cup\{1,4,12,22\}$ that belong to only one of them.
Comparing $\{0,3,11,21,26\}$ with the other 25 translates shows that for each pair there are eight or more elements that are not shared.

Suppose that the receiver would like to locate the boundaries of each frame in a sequence of qutrits.
We assume that the receiver knows that the same quantum tag $T$ is attached to every frame.
As defined in the previous preliminaries subsection,
the receiver obtains a binary outcome vector by performing measurement $\{\bra{0}\ket{0} + \bra{1}\ket{1}, \bra{2}\ket{2}\}$ on each qutrit.
In our scheme, the receiver compares the outcome vector $\boldsymbol{y}$
of each set of $v$ consecutive qutrits with the known correct tag vector $\boldsymbol{x}_F$.
If the quantum channel is noiseless, it suffices to check whether the two vectors are identical.
In fact, if the quantum tag is of comma-free index $\rho \geq 1$,
there are at least $\rho$ discrepancies between the correct tag vector and the outcome vector when there is misalignment.
A similar strategy works for the noisy case if the quantum tag is carefully designed:
\begin{theorem}\label{th1}
Assume that every frame consists of $v$ qutrits and is tagged by the same self-synchronizing quantum tag of length $v$ and comma-free index $\rho$.
The boundaries of each frame can be located if the number of erasures and incursions in any consecutive $v$ qutrits is
at most $\lfloor\frac{\rho-1}{2}\rfloor$.
\end{theorem}
\begin{IEEEproof}
Let $\boldsymbol{x}$ be the tag vector of a self-synchronizing quantum tag of length $v$ and comma-free index $\rho$.
Define $\mathcal{C} = \{\pi^t(\boldsymbol{x}) \ \vert \ 0 \leq t \leq v-1\}$.
Then $\mathcal{C}$ is a classical binary error-correcting code of minimum distance at least $\rho$.
Consider the nearest neighbor decoder $D$ for the classical code
that takes a $v$-dimensional vector $\boldsymbol{y}$ as input
and randomly returns one of the nearest valid codewords in $\mathcal{C}$ in terms of Hamming distance.
If the Hamming distance between $\boldsymbol{y}$ and a nearest valid codeword is less than or equal to $\lfloor\frac{\rho-1}{2}\rfloor$,
the return from $D$ is unique.
Take consecutive $v$ qutrits $G = (q_0,\dots,q_{v-1})$ from a framed sequence of qutrits
in which each frame is of the same size and tagged by $\text{supp}(\boldsymbol{x})$.
Perform the measurement $\{\bra{0}\ket{0} + \bra{1}\ket{1},  \bra{2}\ket{2}\}$ on each qutrit.
Decode the outcome vector by $D$.
The number of erasures and incursions corresponds to the number of bit flips in the outcome vector.
Thus, by assumption, $D$ returns the uniquely determined nearest codeword in $\mathcal{C}$.
If the result of decoding is $\pi^t(\boldsymbol{x})$, $G$ is $t$ qutrits away to the left from proper alignment.
\end{IEEEproof}

As proved above, the return of the classical decoder $D$ serves as the syndrome for frame synchronization.
Hence, not only can the receiver detect misalignment, but also identify its magnitude.

While we employed the nearest neighbor decoder, any general decoder for classical binary error-correcting codes can be used.
If the receiver has knowledge of anomalies of the quantum channel such as asymmetry in the probabilities of the two types of error, 
this can be exploited to achieve more robust frame synchronization than what the statement of Theorem \ref{th1} suggests.

The weight $k$ of a self-synchronizing $(v,k,\rho)$-quantum tag is the number of qutrits we insert for synchronization
while the length $v$ specifies the size of the payload once $k$ is given.
The comma-free index $\rho$ determines how secure frame synchronization can be.
In view of Theorem \ref{th1}, we are interested in self-synchronizing $(v,k,\rho)$-quantum tags that achieve the largest $\rho$ for given $v$ and $k$.
The following theorem gives an upper bound on the comma-free index:
\begin{theorem}\label{ubound1}
If there exists a self-synchronizing $(v,k,\rho)$-quantum tag, then
\[\rho \leq \frac{2k(v-k)}{v-1}.\]
\end{theorem}
\begin{IEEEproof}
Let $\boldsymbol{x}$ be the tag vector of a self-synchronizing $(v,k,\rho)$-quantum tag.
Because the maximum in a finite multiset of integers is larger than or equal to their arithmetic mean, we have
\begin{align*}
\max_{1 \leq t \leq v-1}\left(R_{\boldsymbol{x}}(t)\right) &\geq \frac{1}{v-1}\sum_{1 \leq t \leq v-1}R_{\boldsymbol{x}}(t)\\
&= \frac{k(k-1)}{v-1}.
\end{align*}
Hence, by the definition of comma-free index, we have
\begin{align*}
\rho &\leq 2\left(k - \max_{1 \leq t \leq v-1}\left(R_{\boldsymbol{x}}(t)\right)\right)\\
&\leq 2\left(k - \frac{k(k-1)}{v-1}\right)\\
&= \frac{2k(v-k)}{v-1}.
\end{align*}
The proof is complete.
\end{IEEEproof}

A self-synchronizing $(v,k,\rho)$-quantum tag is \textit{optimal} if $\rho = \lfloor \frac{2k(v-k)}{v-1} \rfloor$.
We employ combinatorial design theory to construct optimal self-synchronizing quantum tags.

A \textit{cyclic difference set}, denoted as cyclic DS$(v,k,\mu)$,
of \textit{order} $v$, \textit{block size} $k$, and \textit{index} $\mu$ is a set $B$ of $k$ non-negative integers less than $v$
such that every element of $\textit{\textbf{Z}}_v \setminus \{0\}$ appears exactly $\mu$ times each
as the difference $a - b \pmod{v}$ between two distinct elements $a, b \in B, a \not= b$.
For instance, take a triple $\{0,1,3\}$.
Then, each nonzero element of $\textit{\textbf{Z}}_7$ appears as the difference between two elements exactly once as follows:
\begin{align*}
1-0 &\equiv 1 \pmod{7}, & 0-1 &\equiv 6 \pmod{7},\\
3-1 &\equiv 2 \pmod{7}, & 1-3 &\equiv 5 \pmod{7},\\
3-0 &\equiv 3 \pmod{7}, & 0-3 &\equiv 4 \pmod{7}.
\end{align*}
Thus, it is a cyclic difference set of order seven, block size three, and index one.
To avoid triviality, we always assume that $k > \mu >0$.
It is known that a cyclic difference set gives a binary sequence with the ideal two-level auto-correlation property (see, for example, \cite{GG}).

Intuitively, an ideal binary sequence would take full advantage of all $1$s it contains at any time shift $t$ extremely efficiently
to give as spiky autocorrelations as possible,
which suggests that the off-peak autocorrelations of such an ideal sequence of given length would be a function of the number of $1$s.
In our problem, the number of $1$s in a sequence represents the information rate penalty we must pay to achieve error-tolerant synchronization.
Therefore, binary sequences with the ideal two-level auto-correlation property would offer the best possible trade-off between
information rates and off-peak autocorrelations.
This heuristic argument is indeed correct;
self-synchronizing quantum tags meeting the bound given in Theorem \ref{ubound1} with equality are equivalent to cyclic difference sets:
\begin{theorem}\label{th2}
A set of integers is an optimal self-synchronizing $(v,k,\rho)$-quantum tag with $\rho = \frac{2k(v-k)}{v-1}$
if and only if it is a cyclic \textup{DS}$(v,k,\mu)$.
\end{theorem}
\begin{IEEEproof}
Take a set of integers of cardinality $k$.
Without loss of generality, we assume that the set is a $k$-subset $T \subseteq \textit{\textbf{Z}}_v$ of the ring of integers modulo $v$.
By the proof of Theorem \ref{ubound1}, $T$ is a self-synchronizing $(v,k,\rho)$-quantum tag with $\rho = \frac{2k(v-k)}{v-1}$ if and only if
the corresponding tag vector $\boldsymbol{x}$ satisfies the condition that
\[\max_{1 \leq t \leq v-1}\left(R_{\boldsymbol{x}}(t)\right) = \frac{1}{v-1}\sum_{1 \leq t \leq v-1}R_{\boldsymbol{x}}(t),\]
which holds if and only if the maximum is equal to the arithmetic average.
Because $R_{\boldsymbol{x}}(t) = \vert T \cap (T+t)\vert $,
this is equivalent to the condition that for any $t \not\equiv 0 \pmod{v}$
\[\vert T \cap (T+t)\vert = \frac{k(k-1)}{v-1}.\]
Because $T$ and $T+t$ share elements if and only if $T$ has elements $a$ and $a+t \pmod{v}$,
the condition holds if and only if $T$ is a cyclic DS$(v,k,\mu)$ with $\mu = \frac{k(k-1)}{v-1}$.
Thus, it suffices to show that every cyclic difference set of order $v$ and block size $k$ is of index $\mu = \frac{k(k-1)}{v-1}$.
In a set of cardinality $k$, there are exactly ${{k}\choose{2}} = \frac{k(k-1)}{2}$ pairs of elements.
Because a cyclic difference set contains every nonzero element of $\textit{\textbf{Z}}_v$ as the difference between elements exactly $\mu$ times,
we have
\[\mu(v-1) = 2\cdot\frac{k(k-1)}{2}.\]
Hence, any cyclic DS$(v,k,\mu)$ satisfies the condition that $\mu = \frac{k(k-1)}{v-1}$.
The proof is complete.
\end{IEEEproof}

With this equivalence, we only need to look for cyclic difference sets
to find optimal self-synchronizing quantum tags that attain the upper bound on the comma-free index with equality.
The study of cyclic difference sets has a long history
with one of the earliest examples in modern mathematics being Kirkman's construction for projective planes from the 19th century \cite{Kirk}.
Here we list useful results on cyclic difference sets that can be described in simple mathematics.
For more complete treatments on the basics of the theory of difference sets, we refer the reader to \cite{BJL}.
For more recent results, a concise survey on recent progress in algebraic design theory can be found in \cite{Xiang}.

The following is a trivial but useful property of a cyclic difference set:
\begin{proposition}
The complement of a cyclic \textup{DS}$(v,k,\mu)$ is a cyclic \textup{DS}$(v,v-k,v-2k+\mu)$.
\end{proposition}
Because index $\mu$ must be an integer, no cyclic DS$(v,\frac{v}{2},\mu)$ exists.
Thus, we only need to consider the case when $k < \frac{v}{2}$.

Projective geometry provides proof of the existence of cyclic difference sets for particular parameters.
To exploit finite geometry, we define special combinatorial designs.
A \textit{simple} $2$-\textit{design} of \textit{order} $v$, \textit{block size} $k$, and \textit{index} $\mu$
is an ordered pair $(V, \mathcal{B})$, where $V$ is a finite set of cardinality $v$
and $\mathcal{B}$ is a set of $k$-subsets of $V$ such that each pair of elements of $V$ is included in exactly $\mu$ elements of $\mathcal{B}$.
The elements of $V$ are called \textit{points} while those of $\mathcal{B}$ are \textit{blocks}.
A simple $2$-design is said to be \textit{symmetric} if $\vert \mathcal{B} \vert = \vert V \vert$.
It is \textit{cyclic} if the cyclic group of order $v$ acts regularly on the points.
A simple way to see a cyclic simple $2$-design of order $v$ is that
there is a way to rename points by using integers between $0$ and $v-1$ such that for any block $B \in \mathcal{B}$
the translated block $B+1$ is also a block of the design.
The Fano plane is the smallest nontrivial example: the set
\[\mathcal{B} = \{\{0,1, 3\} + i \ \vert \ i \in \textit{\textbf{Z}}_7\}\]
of triples forms
a cyclic simple $2$-design of order seven, block size three, and index one.
The point set is the seven integers $\{0,1,2,3,4,5,6\}$.
It is easy to check that each pair appears exactly once
while for any block $B \in \mathcal{B}$ we surely have $B+1 \in \mathcal{B}$.
Because $\vert \mathcal{B} \vert = 7 = \vert \{0,1,2,3,4,5,6\} \vert$, it is symmetric as well.

Symmetric designs and difference sets are closely related, as is suggested by the fact that $\{0,1,3\}$ is a cyclic DS$(7,3,1)$.
In fact, it is readily checked that for any cyclic DS$(v,k,\mu)$ $B$
such that the orbit $\text{\textit{Orb}}_{\textit{\textbf{Z}}_v}(B) = \bigcup_{i \in \textit{\textbf{Z}}_v}\left\{B+i\right\}$ is of length $v$,
the $v$ subsets $\text{\textit{Orb}}_{\textit{\textbf{Z}}_v}(B)$ of $\textit{\textbf{Z}}_v$ form a cyclic simple $2$-design of order $v$ and index $\mu$ that is symmetric.
Conversely, any cyclic simple $2$-design gives a cyclic difference set if it is symmetric.

The \textit{points} of the $m$-dimensional projective geometry PG$(m,q)$ over $\mathbb{F}_q$ are the $1$-dimensional subspaces of $\mathbb{F}_{q^{m+1}}$.
The \textit{hyperplanes} of PG$(m,q)$ are the subspaces of codimension one, namely the $m$-dimensional vector subspaces of $\mathbb{F}_{q^{m+1}}$.
It is well-known that the points and hyperplanes of PG$(m,q)$ form the points and blocks of a simple $2$-design
of order $\frac{q^{m+1}-1}{q-1}$, block size $\frac{q^m-1}{q-1}$, and index $\frac{q^{m-1}-1}{q-1}$,
where a Singer cycle acts regularly on the points and also blocks \cite{Hirschfeld}.
This simple $2$-design is symmetric.
Thus, any representative of its block orbit forms a cyclic DS$(\frac{q^{m+1}-1}{q-1},\frac{q^m-1}{q-1},\frac{q^{m-1}-1}{q-1})$.
\begin{theorem}\label{ds1}
For any prime power $q$ and integer $m \geq 2$
there exists an optimal $(\frac{q^{m+1}-1}{q-1},\frac{q^m-1}{q-1},\frac{2(q^m-q^{m-1})}{q-1})$-quantum tag.
\end{theorem}
\begin{IEEEproof}
There exists a cyclic DS$(v,k,\mu)$ if and only if there exists an optimal $(v,k)$-quantum tag of comma-free index
\begin{align*}
\rho &= \frac{2k(v-k)}{v-1}\\
&= 2\left(k - \mu\right).
\end{align*}
Since a hyperplane of PG$(m,q)$ forms a cyclic difference set,
plugging $v = \frac{q^{m+1}-1}{q-1}$, $k = \frac{q^m-1}{q-1}$, and $\mu = \frac{q^{m-1}-1}{q-1}$ proves the assertion.
\end{IEEEproof}

There are various known explicit constructions which may give nonisomorphic cyclic difference sets of these parameters.
Since we only need one explicit construction, a straightforward method for realizing the parameters given in Theorem \ref{ds1} is to take the set
\[\left\{i \ \middle\vert \ 0 \leq i \leq \frac{q^{m+1}-1}{q-1}, \text{Tr}_{(m+1)/1}(\alpha^i)=0\right\}\]
modulo $\frac{q^{m+1}-1}{q-1}$, where $\alpha$ is a generator of the multiplicative group of $\mathbb{F}_{q^{m+1}}$
and $\text{Tr}$ is the trace function $\text{Tr}_{(m+1)/1}(\beta) = \sum_{i=0}^{m}\beta^{q^i}$ from $\mathbb{F}_{q^{m+1}}$ onto $\mathbb{F}_q$.
The resulting set forms the Singer difference set \cite{Singer}.

Another simple method for explicitly constructing cyclic difference sets is to use the concept of cyclotomy in the theory of finite fields.
The following quantum tags are classical examples of cyclic difference sets obtained through cyclotomy:
\begin{theorem}\label{cd1}
For any prime $p \equiv 3 \pmod{4}$ the quadratic residues $\{x^2 \ \vert \ x \in \mathbb{F}_p\setminus\{0\}\}$
form an optimal $(p, \frac{p-1}{2},\frac{p+1}{2}) $-quantum tag.
\end{theorem}
\begin{theorem}\label{cd2}
For any prime $p = 4t^2+1$ with $t$ odd, the set $\{x^4 \ \vert \ x \in \mathbb{F}_p\setminus\{0\}\}$
forms an optimal $(p, \frac{p-1}{4},\frac{3p+1}{8})$-quantum tag.
\end{theorem}
\begin{theorem}\label{cd3}
For any prime $p = 4t^2+9$ with $t$ odd, the set $\{x^4 \ \vert \ x \in \mathbb{F}_p\}$
forms an optimal $(p, \frac{p+3}{4},\frac{3p+9}{8})$-quantum tag.
\end{theorem}
\begin{theorem}\label{cd4}
For any prime $p = 8t^2+1=64u^2+9$ with $t$ and $u$ both odd, the set $\{x^8 \ \vert \ x \in \mathbb{F}_p\setminus\{0\}\}$
forms an optimal $(p, \frac{p-1}{8},\frac{7p+1}{32})$-quantum tag.
\end{theorem}
\begin{theorem}\label{cd5}
For any prime $p = 8t^2+49=64u^2+441$ with $t$ odd and $u$ even, the set $\{x^8 \ \vert \ x \in \mathbb{F}_p\}$
forms an optimal $(p, \frac{p+7}{8},\frac{7p+49}{32})$-quantum tag.
\end{theorem}
For the proof of the fact that the sets defined in \Cref{cd1,cd2,cd3,cd4,cd5} form cyclic difference sets,
we refer the reader to a standard textbook in this field such as \cite{Storer}.

The following is another series of optimal quantum tags from cyclotomic cyclic difference sets called \textit{Hall difference sets} \cite{Halldif}:
\begin{theorem}
Let $p \equiv 1 \pmod{6}$ be a prime of the form $p = 4t^2+27$ and $\alpha$ a primitive root of $p$ such that
if $3 \equiv \alpha^x \pmod{p}$, then $x \equiv 1 \pmod{6}$.
The set $\{\alpha^i \ \vert \ i \equiv 0, 1, 3 \pmod{6}\}$
forms an optimal $(p,\frac{p-1}{2},\frac{p+1}{2})$-quantum tag.
\end{theorem}

A pair of primes $p$ and $p+2$ are called \textit{twin primes}.
The following is a special case of the well-known twin prime power difference set (see \cite{Baumert}):
\begin{theorem}
Let $p$ and $p+2$ be primes.
The set of ordered pairs $(x,y)$, $x \in \mathbb{F}_p$, $y \in \mathbb{F}_{p+2}$
such that $x, y$ are both nonzero squares or both nonsquares or $y = 0$
forms an optimal $(p^2+2p,\frac{p^2+2p-1}{2},\frac{p^2+2p+1}{2})$-quantum tag.
\end{theorem}
As Baumert puts in his lecture notes \cite{Baumert},
the corresponding difference sets ``seem to belong to that special class of mathematical objects which are prone to independent rediscovery.''
References to several early instances of independent discovery as well as a proof of more general theorem can be found in \cite{Baumert}.

Whether there exist infinitely many twin primes is one of the most elusive open problems in number theory.
The first Hardy-Littlewood conjecture \cite{HL} suggests that the number $n(x)$ of primes $p \leq x$ such that $p+2$ is also a prime is asymptotically equal to
\[n(x) \sim 2\Pi_2\int_2^x\frac{dx}{(\ln x)^2},\]
where
\[\Pi_2 = \prod_{\substack{p>2\\ p\ \text{prime}}}\frac{p(p-2)}{(p-1)^2}.\]
Small twin primes are listed on the website The On-Line Encyclopedia of Integer Sequences \cite{A014574}.

Computer searches are effective for small parameters.
A comprehensive list of known cyclic DS$(v,k,\mu)$s for $k \leq 300$ can be found at Center for Communications Research's website \cite{DSlist}.

\subsection{Orthogonal quantum tags}\label{s3c}
The previous subsection focused on the situation where each frame is tagged by the same pattern of $\bra{2}$s.
This is a reasonable assumption because if we use two or more quantum tags,
we still have to consider the special case in which some consecutive frames happen to be tagged by the same pattern.
Hence, using multiple tags do not offer more robust frame synchronization in a general setting.

However, with the previous method, we need to decode quantum error-correcting codes in order to obtain any metadata beyond boundary information.
If additional metadata are relatively minor, it would make more sense if the receiver could obtain them without decoding protected qubits.

The use of multiple quantum tags makes it possible to directly retrieve minor metadata by locating $\bra{2}$s.
In fact, if a set $\mathcal{S}$ of quantum tags possesses orthogonality in such a way that
the receiver can both locate a quantum tag and identify which,
tagged frames can naturally carry $\vert \mathcal{S} \vert$-ary classical information by regarding each frame as a logical $\vert \mathcal{S} \vert$-ary digit.
This subsection is devoted to a scheme of this type.

As in the previous method, we assume that each frame is of the same size $v$ and contains $k$ $\bra{2}$s.
The difference is that the pattern of inserted $\bra{2}$s may be different from frame to frame.
In order to mathematically define the kind of orthogonality we need,
we take more advantage of classical coding theory.

A \textit{self-synchronizing code} $\mathcal{C} \subset \mathbb{F}_2^v$ is a binary block code of length $v$ where the 
bit string formed by an overlapped portion of any two concatenated codewords is not a valid codeword.
If the channel is assumed to be noiseless,
the property that no codeword appears as a substring of two adjacent codewords allows for block synchronization without external help.

A well-known classical code is typically not self-synchronizing.
For instance, a linear code of length $v$ must have the $v$-dimensional zero vector as its codeword.
Hence, if the information source transmits two zero vectors in a row,
a bit string of length $v$ that strides the two is still a run of zeros, which is a valid codeword.
Self-synchronizing codes are those that avoid such occurrences of valid codewords when there is misalignment.

We consider an error-tolerant variant of self-synchronizing codes.
Take a sequence $S$ of codewords of a binary block code $\mathcal{C} \subseteq\mathbb{F}_2^v$ of length $v$.
A {\it splice} of length $v$ between codeword $\boldsymbol{c} = (c_0, c_1, \dots, c_{v-1})$
and the following codeword $\boldsymbol{c}' = (c'_0, c'_1, \dots, c'_{v-1})$ in $S$ is
a concatenated binary sequence $(c_{v-i}, \dots, c_{v-1}, c'_0, \dots c'_{v-i-1})$ composed of the last $i$ bits of $\boldsymbol{c}$
and the first $v-i$ bits of $\boldsymbol{c}'$ for some positive integer $i \leq v-1$.
A binary block code of length $v$ is said to be of \textit{comma-free index} $\rho_c$ if the Hamming distance between any codeword $\boldsymbol{c}''$
and any splice of length $v$ between any two codewords $\boldsymbol{c}, \boldsymbol{c}'$ is at least $\rho_c$.
Informally, a code of length $v$ and comma-free index $\rho_c$ has the property that
there are at least $\rho_c$ discrepancies between a properly aligned valid codeword and a $v$ bit substring of two adjacent codewords when compared bit by bit.

Comma-free indices of classical binary codes are similar to minimum distances in that
a self-synchronizing code of length $v$ and comma-free index $\rho_c$
assures block synchronization under the presence of up to $\lfloor\frac{\rho_c-1}{2}\rfloor$ bit flips in the received bit sequence of length $v$.
In fact, to achieve synchronization, the receiver only needs to compare the received consecutive $v$ bits
with the valid codewords and look up the nearest neighbors.
If there are less than or equal to $\lfloor\frac{\rho_c-1}{2}\rfloor$ bits that defer between the consecutive $v$ bits and some valid codeword $\boldsymbol{c}$,
assuming that the channel does not produce more than $\lfloor\frac{\rho_c-1}{2}\rfloor$ bit flips,
the receiver can safely declare that the string of bits at hand are properly aligned.
If there is no codeword within $\lfloor\frac{\rho_c-1}{2}\rfloor$ bits in terms of Hamming distance, there must be misalignment.
As in the minimum distance decoding, if the channel produces more than $\lfloor\frac{\rho_c-1}{2}\rfloor$ bit flips,
the receiver may wrongly assume that the consecutive $v$ bits are properly aligned when there is misalignement.
In this sense, the comma-free index of a block code determines its block synchronization capability.

As is apparent from the above argument,
the tag vector of a self-synchronizing $(v,k,\rho)$-quantum tag may be regarded as a self-synchronizing code of comma-free index $\rho$ with only one codeword.
With this viewpoint, if the tag vectors of frames are chosen so that they from a self-synchronizing code $\mathcal{S}$,
each frame can be seen as a codeword of $\mathcal{S}$ that carries classical $\vert \mathcal{S} \vert$-ary information.
The goal of this subsection is to define and construct special patterns of $\bra{2}$s that form a self-synchronizing code with a large number of codewords
while keeping in mind the special requirement that we are not allowed to insert too many $\bra{2}$s in a frame.

To correctly interpret classical metadata attached as codewords of a self-synchronizing code,
we need to both locate and identify codewords.
As is well-known, a binary block code of minimum distance $d$ can identify codewords
if synchronization is correct and the number of bit flips in a received vector is up to $\lfloor \frac{d-1}{2}\rfloor$.

We use the supports of binary self-synchronizing codes as quantum tags.
A set $\mathcal{S}$ of self-synchronizing quantum tags is of \textit{comma-free index} $\rho_c$ and of \textit{distance} $d$
if the set of corresponding tag vectors forms a self-synchronizing code of comma-free index $\rho_c$ and minimum distance at least $d$.
We say that the elements of $\mathcal{S}$ are \textit{orthogonal} self-synchronizing $(v,k,\rho_c,d;\vert\mathcal{S}\vert)$-quantum tags
if comma-free index $\rho_c$ and distance $d$ are both positive
and for any $S \in \mathcal{S}$, the cardinality $\vert S \vert = k$.
With this definition, the positions of $1$s in a codeword of a self-synchronizing code specify where $\bra{2}$s are inserted in a frame
tagged by the corresponding quantum tag.
The $0$s in a codeword represent the payload qubits protected by a quantum error-correcting code.

The following proposition summarizes the argument given above in this subsection:
\begin{proposition}
Assume that every frame consists of $v$ qutrits and is tagged by one of orthogonal self-synchronizing $(v,k,\rho_c,d;s)$-quantum tags.
The location of the boundaries of each frame and additional metadata in the form of one classical $s$-ary digit
can be obtained if the number of erasures and incursions in any consecutive $v$ qutrits is
at most $\lfloor\frac{\min(\rho_c, d)-1}{2}\rfloor$.
\end{proposition}

Self-synchronizing codes and related concepts have been investigated in a variety of fields from various viewpoints.
Noticeable progress has been made on the theoretical side in recent research (see, for example, \cite{Polyanskiy,HighrateSC}).
Unfortunately, it appears that the known self-synchronizing codes are generally unsuitable for our purpose
due to the fact that known classical self-synchronizing codes are typically cosets of linear codes and are thus of variable weight.
In fact, while various self-synchronizing codes that are of large alphabet size and contain codewords with many $1$s are available,
the binary case with emphasis on small constant weight $k$ does not seem to have received much attention in the literature.
Making things more difficult, whether self-synchronizing or not, finding a good code of small constant weight is already a difficult problem on its own.
Nonetheless, as will be shown in the remainder of this subsection,
there exist infinitely many nontrivial self-synchronizing error-correcting codes of small constant weight.

Recall that the symbol $\pi$ refers to the cyclic shift of a vector by one bit to the right.
The \textit{periodic cross-correlation} $R_{\boldsymbol{x},\boldsymbol{y}}(t)$ at \textit{time} $t$ of $v$-dimensional vectors
$\boldsymbol{x} = (x_0, \dots, x_{v-1})$ and $\boldsymbol{y} = (y_0, \dots, y_{v-1})$ over $\mathbb{F}_2$ is defined as the inner product
\begin{align*}
R_{\boldsymbol{x},\boldsymbol{y}}(t) &= \boldsymbol{x}\cdot\pi^t(\boldsymbol{y})\\
&= \sum_{0 \leq i \leq v-1}x_iy_{i+t}
\end{align*}
between $\boldsymbol{x}$ and a cyclic shift of $\boldsymbol{y}$, where the subscripts in the sum are taken modulo $v$.
The periodic autocorrelations we defined in the previous subsection are the cross-correlations of $\boldsymbol{x}$ and itself.

A $(v,k,\lambda_a,\lambda_c)$ \textit{optical orthogonal code}
$\mathcal{C} \subseteq \mathbb{F}_2^v$ of \textit{length} $v$ and \textit{weight} $k$ is
a set of $v$-dimensional binary vectors of weight $k$ such that
for any $\boldsymbol{c} \in \mathcal{C}$ its off-peak periodic autocorrelations are at most $\lambda_a$
and for any pair of distinct codewords $\boldsymbol{c}, \boldsymbol{c}' \in \mathcal{C}$ their periodic cross-correlations are at most $\lambda_c$.
In other words, it is a set of $v$-dimensional vectors with $k$ $1$s and $v-k$ $0$s whose coordinates are indexed by $\textit{\textbf{Z}}_v$ such that
\[R_{\boldsymbol{c}}(t) = \sum_{0 \leq i \leq v-1}c_i c_{i+t} \leq \lambda_a\]
for any $\boldsymbol{c} = (c_0, c_1, \dots, c_{v-1}) \in \mathcal{C}$ and any nonzero element $t \in \textit{\textbf{Z}}_v$ and such that
\[R_{\boldsymbol{c}, \boldsymbol{c'}}(t) = \sum_{0 \leq i \leq v-1}c_i c'_{i+t} \leq \lambda_c\]
for any pair of distinct vectors $\boldsymbol{c} = (c_0, c_1, \dots, c_{v-1}),  \boldsymbol{c}' = (c'_0, c'_1, \dots, c'_{v-1})\in \mathcal{C}$
and any $t \in \textit{\textbf{Z}}_v$.
The original motivation of the study of optical orthogonal codes was to realize
code-division multiple-access (CDMA) fiber optical communications \cite{CSW},
where asynchronous location and identification of codewords is important.
For codeword location and identification, codes with a very high degree of orthogonality are desirable \cite{FD},
which translates to the condition that off-peak autocorrelations and cross-correlations should be as small as possible.
To keep the subsequent discussion simple,
we only consider the case when $\lambda_a = \lambda_c = 1$.
An optical orthogonal code with this restriction is said to be of \textit{index} one and will be denoted as $(v,k,1)$-OOC.
To avoid triviality, we also assume that $k \geq 2$.

\begin{theorem}\label{ooc}
The supports of the codewords of a $(v,k,1)$-\textup{OOC} $\mathcal{C}$ are orthogonal self-synchronizing $(v,k,k-2,2k-2; \vert\mathcal{C}\vert)$-quantum tags.
\end{theorem}
\begin{IEEEproof}
It suffices to show that a $(v,k,1)$-OOC $\mathcal{C}$ is of comma-free index $k-2$ and minimum distance at least $2k-2$.
By definition, for any two distinct codewords $\boldsymbol{c}, \boldsymbol{c}' \in \mathcal{C}$,
the cross-correlation $C_{\boldsymbol{c}, \boldsymbol{c}'}(0)$ at time $0$ is at most one.
Thus, the Hamming distance between any pair of distinct codewords is at least $2k - 2$.
To prove that the code is of comma-free index $k-2$, consider a splice of a pair of codewords.
Take a codeword $\boldsymbol{c} \in \mathcal{C}$.
Because off-peak autocorrelations and cross-correlations are both at most one,
there are at most two coordinates at which both $\boldsymbol{c}$ and a splice have $1$s.
Because any codeword in $\mathcal{C}$ has $k$ $1$s, the Hamming distance between a valid codeword and a splice is at least $k-2$.
Thus, $\mathcal{C}$ is of comma-free index $k-2$.
\end{IEEEproof}

We would like optical orthogonal codes with the largest possible number of codewords for given $v$ and $k$.
The Johnson bound gives the upper bound on the number of codewords \cite{Johnson}:
\begin{theorem}\label{jb}
Let $\mathcal{C}$ be a $(v,k,1)$-\textup{OOC}.
Then it holds that
\[
\vert \mathcal{C} \vert \leq
\left\lfloor\frac{1}{k}\left\lfloor\frac{v-1}{k-1}\right\rfloor\right\rfloor.
\]
\end{theorem}

A $(v,k,1)$-OOC is \textit{optimal} if the number of codewords attains this upper bound.
When $v-1$ is divisible by $k$ and $k-1$, the right-hand side of the Johnson bound does not suffer from the slight drop due to the floor functions.
Table \ref{ooctable} lists well-known classes of optimal optical orthogonal codes that have exactly $\frac{v-1}{k(k-1)}$ codewords.
\begin{table*}
\renewcommand{\arraystretch}{1.6}
\caption{Optimal optical orthogonal codes}
\label{ooctable}
\centering
\begin{tabular}{ccccc}
\hline\hline
\bfseries Length $v$ &\bfseries Weight $k$ &\bfseries Number $\vert \mathcal{C} \vert$ of codewords &\bfseries Constraint &\bfseries Reference\\
\hline
\multirow{3}{*}{$p$} & \multirow{3}{*}{$k$} & \multirow{3}{*}{$\frac{p-1}{k(k-1)}$} &
$p \equiv 1 \pmod{k(k-1)}$ prime, & \multirow{3}{*}{\cite{WilsonC}}\\
 & & & $p > c_k$, &\\
 & & & $c_k$ constant dependent on $k$\rlap{\textsuperscript{a}} &\\
$q^t-1$ & $q$ & $\frac{q^{t-1}-1}{q-1}$ & $q$ prime power&
Affine geometry with origine deleted\rlap{\textsuperscript{b, c}}\\
$\frac{q^{t+1}-1}{q-1}$ & $q+1$ & $\begin{cases}\frac{q^t-1}{q^2-1}, \ t\ \text{even}\\ \frac{q^t-1}{q^2-1}, \ t\ \text{odd}\end{cases}$ &
$q$ prime power& Projective geometry\rlap{\textsuperscript{d}}\\
\hline\hline
\multicolumn{5}{l}{\scriptsize\textsuperscript{a}
$c_5 = 0$ \cite{CZ}. $c_6 = 61$ \cite{CZ2}.
For $k \geq 7$ in general, a sufficient condition is prime $p > c_k = {{k}\choose{2}}^{k(k-1)}$ that is congruent to one modulo $k(k-1)$.}\vspace{-1.1mm}\\
\multicolumn{5}{l}{\scriptsize\textsuperscript{b}
This is also known as Euclidean geometry in classical and quantum coding theory \cite{KLF,FCVBT}.}\vspace{-1.1mm}\\
\multicolumn{5}{l}{\scriptsize\textsuperscript{c}
The same parameters may be realized as a generalized Bose-Chowla family \cite{MOKL}.}\vspace{-1.1mm}\\
\multicolumn{5}{l}{\scriptsize\textsuperscript{d}
For a proof, see a standard reference book such as \cite{Hirschfeld}.}\vspace{3.3mm}
\end{tabular}
\end{table*}
For instance, the following set of orthogonal self-synchronizing quantum tags is the well-known cyclotomic OOC \cite{WilsonC}:
\begin{theorem}
There exists a constant $c_k$ that depends on $k$ such that for any prime $p \equiv 1 \pmod{k(k-1)}$ greater than $c_k$
there exists a set of orthogonal self-synchronizing $(p,k,k-2,2k-2;\frac{p-1}{k(k-1)})$-quantum tags.
\end{theorem}
There are numerous other constructions and existence results on optical orthogonal codes.
The most recent results can be found in \cite{HandbookCD,Momi,YYL} and references therein.

Note that optical orthogonal codes have been studied under various different names.
Optimal $(v,k,1)$-OOCs with $v-1$ divisible by $k$ and $k-1$ are also known as \textit{cyclic difference families}.
Such an OOC is also equivalent to a system of representatives of block orbits of a cyclic simple $2$-design of order $v$, block size $k$ and index one.
When the divisibility of $v-1$ is not of concern, they may be called \textit{cyclic difference packings without short orbits}.
The equivalence between these mathematical objects was proved in \cite{FM}.
The definitions of cyclic difference families and packings as well as known facts and various other related concepts can be found in \cite{HandbookCD}.

\subsection{Quantum packet headers}\label{s3d}
In this final subsection of Section \ref{framework},
we briefly discuss how to use our quantum tags when the number of qubits in the payload may be different from frame to frame.
Such generalization may be more natural and reasonable in some situations.
For instance, it makes sense to protect important quantum information by a long, strong quantum error-correcting code
while using a shorter one for less important quantum information to optimize the communication rate.

Quantum tags can be exploited when variable length frames are allowed as well.
The key idea is to mimic the concept of packet headers.

In classical communications networks, overhead data are often attached at the beginning or the end of a frame as extra bits to signal the boundaries.
In the simplest method we introduced in Section \ref{model}, single $\bra{2}$s are inserted to indicate the beginnings of frames.
Because self-synchronizing tags are locatable even under the presence of quantum errors,
we can use a very short one in place of the single $\bra{2}$s.
In fact, by following the same type of argument as in the previous two subsections,
one can easily prove that a DS$(v,k,\mu)$ and a $(v,k,1)$-OOC give $v$ qutrit blocks that are locatable under the presence of up to
any $\lfloor\frac{k-\mu-1}{2}\rfloor$ and $\lfloor\frac{k-2}{2}\rfloor$ quantum errors respectively.
By the same token, it is also straightforward to see that the scheme based on a DS$(v,k,\mu)$ and a $(v,k,1)$-OOC tolerates
up to $k-\mu-1$ and $k-2$ erasures if they are the only possible kind of error.

To illustrate this method,
assume that we would like to transmit qubits over the quantum erasure channel with one supplementary qubit in each frame as metadata.
We encode one supplementary qubit by the four qubit code \cite{GBP}.
Take a DS$(7,3,1)$, which forms a self-synchronizing $(7,3,4)$-quantum tag.
This tag makes the encoded physical four qubits locatable under the presence of up to one erasure on any qutrit.
Hence, the resulting seven qutrits can be seen as a stand-alone locatable block
that holds one qubit of information and can correct up to one erasure such as photon loss during optical quantum information processing \cite{KLM,OFV}.
We attach this short block to the beginning of a long payload (see Fig.\ \ref{fig2}).
\setlength{\unitlength}{8.4mm}
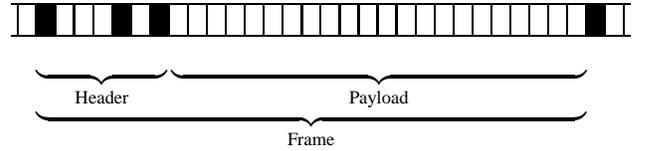
\begin{figure}
\centering
\begin{picture}(10,0.5)
\put(0,0){\line(1,0){9.8}}
\put(0,0.5){\line(1,0){9.8}}
\put(0.1,0){\line(0,1){0.5}}
\put(0.4,0){\line(0,1){0.5}}
\put(0.7,0){\line(0,1){0.5}}
\put(1.0,0){\line(0,1){0.5}}
\put(1.3,0){\line(0,1){0.5}}
\put(1.6,0){\line(0,1){0.5}}
\put(1.9,0){\line(0,1){0.5}}
\put(2.2,0){\line(0,1){0.5}}
\put(2.5,0){\line(0,1){0.5}}
\put(2.8,0){\line(0,1){0.5}}
\put(3.1,0){\line(0,1){0.5}}
\put(3.4,0){\line(0,1){0.5}}
\put(3.7,0){\line(0,1){0.5}}
\put(4,0){\line(0,1){0.5}}
\put(4.3,0){\line(0,1){0.5}}
\put(4.6,0){\line(0,1){0.5}}
\put(4.9,0){\line(0,1){0.5}}
\put(5.2,0){\line(0,1){0.5}}
\put(5.5,0){\line(0,1){0.5}}
\put(5.8,0){\line(0,1){0.5}}
\put(6.1,0){\line(0,1){0.5}}
\put(6.4,0){\line(0,1){0.5}}
\put(6.7,0){\line(0,1){0.5}}
\put(7,0){\line(0,1){0.5}}
\put(7.3,0){\line(0,1){0.5}}
\put(7.6,0){\line(0,1){0.5}}
\put(7.9,0){\line(0,1){0.5}}
\put(8.2,0){\line(0,1){0.5}}
\put(8.5,0){\line(0,1){0.5}}
\put(8.8,0){\line(0,1){0.5}}
\put(9.1,0){\line(0,1){0.5}}
\put(9.4,0){\line(0,1){0.5}}
\put(9.7,0){\line(0,1){0.5}}
\put(0.4,0){\rule{2.52mm}{4.2mm}}
\put(1.6,0){\rule{2.52mm}{4.2mm}}
\put(2.2,0){\rule{2.52mm}{4.2mm}}
\put(9.1,0){\rule{2.52mm}{4.2mm}}
\end{picture}
$\underbrace{\underbrace{\phantom{llllllllllllll0}}_\text{Header}
\underbrace{\phantom{llllllllllllllllllllllllllllllllllllllll0000lll}}_\text{Payload}}_\text{Frame}$ \phantom{lll}
\caption{Frame with a self-synchronizing quantum tag at the beginning. $\vert$
The black boxes are $\bra{2}$s. The white boxes are qubits.
The first seven qutrits in the frame are the ``quantum packet header'' tagged by $\{0,4,6\}$, which is a cyclic DS$(7,3,1)$.
The next frame on the right shown partially is also tagged.
The number of qubits the payload contains is arbitrary and may be different from frame to frame.}\label{fig2}
\end{figure}
Then, the payload together with the seven qutrits form a frame containing an extra short portion
that marks the start of the information block and also carries one qubit of supplementary metadata.

We may call a locatable sequence of qutrits a \textit{quantum packet header} if they are followed by
the main information block consisting of an arbitrary number of qubits in a frame.
We can use any self-synchronizing quantum tag as a quantum packet header.
Trivially, orthogonal self-synchronizing quantum tags can be used the same way to add classical information
in such a way that the receiver does not need to decode a quantum error-correcting code to obtain simple metadata.
The qubits within a quantum tag may also serve as part of the payload if there is no need to add metadata except for boundary locations.

While we can use single and orthogonal quantum tags as suboptimal solutions,
it is unlikely that they give the best possible quantum packet headers.
To formally study quantum packet headers, we need a notion analogous to periodic auto- and cross-correlations. 

The \textit{aperiodic autocorrelation} $C_{\boldsymbol{x}}(t)$ at \textit{time} $t$ of a $v$-dimensional vector
$\boldsymbol{x} = (x_0, \dots, x_{v-1})$ over $\mathbb{F}_2$ is defined by
\[C_{\boldsymbol{x}}(t) = \sum_{0 \leq i \leq v-1}x_ix_{i+t},\]
where $x_i = 0$ for $i$ not in the range $0 \leq i \leq v-1$.
Similarly, the \textit{aperiodic cross-correlation} $C_{\boldsymbol{x},\boldsymbol{y}}(t)$ at \textit{time} $t$ of $v$-dimensional vectors
$\boldsymbol{x} = (x_0, \dots, x_{v-1})$ and $\boldsymbol{y} = (y_0, \dots, y_{v-1})$ over $\mathbb{F}_2$ is defined by
\[C_{\boldsymbol{x},\boldsymbol{y}}(t) = \sum_{0 \leq i \leq v-1}x_iy_{i+t},\]
where $x_i = y_i = 0$ for $i$ not in the range $0 \leq i \leq v-1$.

Single quantum tags need binary sequences with the lowest possible off-peak periodic autocorrelations for a given length and weight.
By the same token, it is straightforward to see that if the payload of each frame consists of a sufficiently large number of qubits,
ideal quantum packet headers are equivalent to binary sequences
with the lowest possible aperiodic autocorrelations $C_{\boldsymbol{x}}(t)$ for $t \not= 0$ for a given length and weight.
Orthogonal quantum packet headers may be defined as a set of quantum tags of low aperiodic cross-correlations and large distance
as in orthogonal quantum tags.

It is widely known in sequence design that aperiodic correlation properties are extremely difficult to analyze.
In fact, in spite of their various important applications, there seem to exist only a few analytical results in the literature.
As far as the author is aware, a theoretical study of aperiodic correlations of binary sequences of fixed weight is nonexistent
except our naive observation that self-synchronizing codes of constant weight give a suboptimal solution.
For known results on designing binary sequences with good aperiodic correlation properties,
we refer the reader to \cite{FD}.

As we have seen,
it is indeed possible to frame a sequence of qubits into units and attach supplementary data by mimicking classical packet framing.
We conclude this section by posing two questions:
\begin{problem}
How can we characterize and construct a binary sequence of the lowest possible aperiodic autocorrelations for a given length and weight?
\end{problem}
\begin{problem}
How can we characterize and construct a constant-weight code of large minimum distance with a large number of codewords
whose binary vector representations have good aperiodic cross-correlations?
\end{problem}

\section{Concluding remarks}\label{conclude}
We developed a coding theoretic method for parsing quantum information
under the presence of quantum noise without assuming prior frame synchronization between the information source and sink.
The notion of asynchronous quantum information processing we introduced opens paths to various possibilities
such as helping local quantum processing involving quantum memory.
In quantum communication scenarios, it may serve as a framework for quantum networks \cite{SO,QnetworkNature}
where users are self-synchronizing and can also attach labels to their quantum information.
Attached metadata are not limited to classical information.
They may be quantum information used as control qubits in a quantum network
or ebits for entanglement-assisted quantum error correction \cite{BDH} in which a few protected qubits play the role of ebits \cite{HYH2,LB}.

While we have made the first crucial step toward realizing frame-wise asynchronous quantum information processing,
as is mentioned in Section \ref{model},
our error model does not consider some specific synchronization errors.
For instance, deletion is one of the most extensively studied types of synchronization error in the classical domain.
Informally, in a communication scenario, deletion represents the kind of situation where the receiver does not realize that a certain bit did not arrive.
For example, assume that the sender uses the repetition code of length three
and transmits the two codewords $000$ and $111$.
Assume also that the second bit of the first frame, say, $000$ did not reach the destination.
Suppose that the next frame, say, $111$ arrived successfully.
If there is no mechanism to detect the loss of the third bit, the first received message will be $001$.
If the receiver simply applies the majority vote decoding,
the received vector $001$ will be accepted as a valid codeword with a bit flip on the third bit.
The receiver realizes that something does not add up only at the very end of the communication,
where, to the eye of the receiver, the last bit of the final codeword never arrives.

Correcting deletion is known to be a very difficult problem in classical coding theory \cite{MBT}.
It would be even more difficult to correct such errors in the quantum domain.
In fact, it appears that our scheme in its current form can not provide protection against this type of error in general.
It would be of interest to study how one can overcome deletion of qubits as well.

Another interesting direction of research is the optimality of orthogonal self-synchronizing quantum tags.
We derived the theoretical upper bound on the robustness of frame synchronization for the case when the same single quantum tag is used.
We also proved that cyclic difference sets are equivalent to optimal self-synchronizing quantum tags.
However, our knowledge on parameters of orthogonal self-synchronizing quantum tags is quite limited.
It would be of importance to understand the relations between the number of orthogonal tags, length, weight, comma-free index, and distance.
It would also be intriguing to provide optimal examples that meet a nontrivial bound that sharply captures a trade-off between parameters.

An equally interesting and challenging problem is to understand the structure of best possible quantum packet headers.
If we take an approach similar to the case of quantum tags by regarding them as binary codes of constant weight,
what we need to do would be generalize self-synchronizing codes to the aperiodic setting.
A theoretical study of such aperiodically self-synchronizing codes of constant weight would be intriguing
from the viewpoints of both classical and quantum information theory.

This paper made the first attempt at developing a theory of frame synchronization in the quantum domain.
We hope that the concepts we introduced will help develop fundamental building blocks
for complex quantum computing and quantum communication.

\section*{Acknowledgment}
The author thanks M.M. Wilde, D. Clark, V.D. Tonchev, and R.M. Wilson for stimulating discussions and valuable comments.
He is grateful to the anonymous reviewers for their helpful comments and suggestions.
The remark that the presented approach is essentially classical preprocessing is due to one of the reviewers.


\begin{IEEEbiographynophoto}{Yuichiro Fujiwara}
(M'10) received the B.S. and M.S. degrees in mathematics from Keio University, Japan,
and the Ph.D. degree in information science from Nagoya University, Japan.

He was a JSPS postdoctoral research fellow with the Graduate School of
System and Information Engineering, Tsukuba University, Japan, and a visiting
scholar with the Department of Mathematical Sciences, Michigan Technological University.
He is currently with the Division of Physics, Mathematics and Astronomy,
California Institute of Technology, Pasadena, where he works as a JSPS postdoctoral research fellow.

Dr.\ Fujiwara's research interests include combinatorics and its interaction with computer science, quantum information science, and electrical engineering,
with particular emphasis on combinatorial design theory, algebraic coding theory, and quantum information theory.
\end{IEEEbiographynophoto}

\end{document}